\documentclass[a4paper,10pt,twoside]{cfc}
\usepackage{graphicx}
\usepackage{amssymb,bm,mathrsfs,bbm,amscd}
\usepackage[tbtags]{amsmath}
\usepackage{lastpage}

\begin{document}
\newtheorem{theorem}{Theorem}
\newtheorem{defn}[theorem]{\bf{Definition}}

\title{A Fractional Lie Group Method For Anomalous Diffusion Equations}
\author{Guo-cheng Wu\footnote{Corresponding
author, E-mail:~wuguocheng2002@yahoo.com.cn. (G.C. Wu)}
\vspace{4mm}\\
Modern Textile Institute, Donghua University, 1882 Yan'an Xilu Road, \\
{Shanghai 200051, China}\\ [6pt] Received 20 May 2010; accepted 13
July 2010}

  \date{}
\maketitle

\begin{abstract}The Lie group method provides an efficient tool
to solve nonlinear partial differential equations. This paper
suggests a fractional partner for fractional partial differential
equations. A space-time fractional diffusion equation is used as an
example to illustrate the effectiveness of the Lie group method.
\end{abstract}


 \setcounter{page}{27}

\section{Introduction}

In the last three decades, researchers have found fractional
differential equations (FDEs) useful in various fields: rheology,
quantitative biology, electrochemistry, scattering theory,
diffusion, transport theory, probability potential theory and
elasticity [1], for details, see the monographs of Kilbas et al.
[2], Kiryakova [3], Lakshmikantham and Vatsala [4], Miller and Ross
[5], and Podlubny [6]. On the other hand, finding accurate and
efficient methods for solving FDEs has been an active research
undertaking.

Since Sophus Lie's work on group analysis, more than 100 years ago,
Lie group theory has become more and more pervasive in its influence
on other mathematical disciplines [7, 8]. Then a question may
naturally arise: is there a fractional Lie group method for
fractional differential equations?

Up to now, only a few works can be found in the literature. For
example, Buckwarand and Luchko derived scaling transformations [9]
for the fractional diffusion equation in Riemann-Liouville sense
\begin{equation}
\frac{{\partial ^\alpha  u(x,t)}}{{\partial t^\alpha  }} =
D\frac{{\partial ^2 \mathop u\limits^{} (x,t)}}{{\partial x^2
}},\;\;0 < \alpha ,\;0 < x{\rm{,}}\;0 < t,\;0 < D. \label{eq1}
\end{equation}

Gazizov et al. find symmetry properties of fractional diffusion
equations of Caputo derivative [10]

\begin{equation}
\frac{{\partial ^\alpha  u(x,t)}}{{\partial t^\alpha  }} =
k\frac{{\partial (\mathop {k(u)u_x }\limits^{} (x,t))}}{{\partial
x}},\;\;0 < \alpha ,\;0 < x{\rm{,}}\;0 < t,\;0 < k. \label{eq2}
\end{equation}

Djordjevic and Atanackovic [11] obtained some similarity solutions
for the time-fractional heat diffusion
\begin{equation}
\frac{{\partial ^\alpha  T(x,t)}}{{\partial t^\alpha  }} =
k\frac{{\partial^{2} (T(x,t))}}{{\partial x^{2}}},\;\;0 < \alpha
,\;0 < x{\rm{,}}\;0 < t. \label{eq3}
\end{equation}

In this study, we investigate anonymous diffusion [12]
\begin{equation}
\frac{{\partial ^\alpha  u(x,t)}}{{\partial t^\alpha  }} =
\frac{{\partial ^{2\beta } u(x,t)}}{{\partial x^{2\beta } }},0 <
\alpha ,\;\beta  \le 1,\;0 < x{\rm{,}}\;0 < t, \label{eq4}
\end{equation}
with a fractional Lie group method, and derive its classification of
solutions. Here the fractional derivative is in the modified
Riemann-Liouville sense [13] and $\frac{{\partial ^{2\beta }
u(x,t)}}{{\partial x^{2\beta } }}$ is defined by$\frac{{\partial
^\beta }}{{\partial x^\beta }}(\frac{{\partial ^\beta
u(x,t)}}{{\partial x^\beta  }}).$

\section{Characteristic Method for Fractional Differential Equations}

Through this paper, we adopt the fractional derivative in modified
Riemann-Liouville sense [13]. Firstly, we introduce some properties
of the fractional calculus that we will use in this study.

(I) Integration with respect to $(dx)^\alpha $(Lemma\textbf{ 2.1} of
[14])

\begin{equation}
_0 I_x^\alpha  f(x) = \frac{1}{{\Gamma (\alpha )}}\int_0^x  (x - \xi
)^{\alpha  - 1} f(\xi )d\xi  = \frac{1}{{\Gamma (\alpha  +
1)}}\int_0^x
 f(\xi )(d\xi )^\alpha  ,0 < \alpha  \le 1.
\label{eq5}
\end{equation}

(II) Some other useful formulas

\begin{equation}
f([x(t)])^{(\alpha )}  = \frac{{df}}{{dx}}x^{(\alpha )} (t), ~\\
{}_0D_x^\alpha  x^\beta   = \frac{{\Gamma (1 + \beta )}}{{\Gamma (1
+\beta  - \alpha )}}x^{\beta  - \alpha } . \\
\label{eq6}
\end{equation}

The properties of Jumarie's derivative were summarized in [13]. The
extension of Jumaire's fractional derivative and integral to
variational approach of several variables is done by Almeida et al.
[15]. Fractional variational interactional method is proposed for
fractional differential equations [16].

It is well known that the method of characteristics has played a
very important role in mathematical physics. Preciously, the method
of characteristics is used to solve the initial value problem for
general first order. With the modified Riemann-Liouville derivative,
Jumaire ever gave a Lagrange characteristic method [17]. We present
a more generalized fractional method of characteristics and use it
to solve linear fractional partial equations.

Consider the following first order equation,
\begin{equation}
a(x,t)\frac{{\partial u(x,t)}}{{\partial x}} + b(x,t)\frac{{\partial
u(x,t)}}{{\partial t}} = c(x,t). \label{eq7}
\end{equation}

The goal of the method of characteristics is to change coordinates
from ${\rm{(}}x,\;t{\rm{)}}$ to a new coordinate system ${\rm{(}}x_0
,\;s{\rm{)}}$ in which the PDE becomes an ordinary differential
equation along certain curves in the $x - t$ plane. The curves are
called the characteristic curves. More generally, we consider to
extend this method to linear space-time fractional differential
equations
\begin{equation}
a(x,t)\frac{{\partial ^{^\beta  } u(x,t)}}{{\partial x^\beta  }} +
b(x,t)\frac{{\partial ^\alpha  u(x,t)}}{{\partial t^\alpha  }} =
c(x,t),0 < \alpha ,\beta  \le 1. \label{eq8}
\end{equation}

With the fractional Taylor's series in two variables [13]

\begin{equation}
du = \frac{{\partial ^{^\beta  } u(x,t)}}{{\Gamma (1 + \beta
)\partial x^\beta  }}(dx)^{^\beta  }  + \frac{{\partial ^\alpha
u(x,t)}}{{\Gamma (1 + \alpha )\partial t^\alpha  }}(dt)^\alpha
,\;\;0 < \alpha ,\;\beta
 \le 1.
\label{eq9}
\end{equation}

Similarly, we derive the generalized characteristic curves
\begin{equation}
\frac{{du}}{{ds}} = c(x,t),
\end{equation}
 \begin{equation}
\frac{{(dx)^{^\beta  } }}{{\Gamma (1 + \beta )ds}} =
a(x,t),\label{eq10}
\end{equation}
\begin{equation}
\frac{{(dt)^\alpha  }}{{\Gamma (1 + \alpha )ds}} = b(x,t). \\
\\
\end{equation}
Eqs. (10)-(12) can be reduced to Jumarie's result if $\alpha  =
\beta $ in [17].

As an example, we consider the fractional equation

\begin{equation}
\frac{{x^\beta  }}{{\Gamma (1 + \beta )}}\frac{{\partial ^\beta
u(x,t)}}{{\partial x^\beta  }} + \frac{{2t^\alpha  }}{{\Gamma (1 +
\alpha )}}\frac{{\partial ^\alpha  u(x,t)}}{{\partial t^\alpha  }} =
0,\;\;0 < \alpha ,\;\beta  \le 1. \label{eq11}
\end{equation}

We can have the fractional scaling transformation

\begin{equation}
u = u(\frac{{x^{^{2\beta } } }}{{\Gamma ^2 (1 + \beta
)}}/\frac{{2t^\alpha  }}{{\Gamma (1 + \alpha )}}). \label{eq12}
\end{equation}

Note that when \begin{math}\alpha  = \beta  = 1{\rm{,}}\end{math} as
is well known, \begin{math}\frac{{x^{^2 } }}{{2t}}\end{math} is one
invariant of the line differential equation

\begin{equation}
x\frac{{\partial u(x,t)}}{{\partial x}} + 2t\frac{{\partial
u(x,t)}}{{\partial t}} = 0. \label{eq13}
\end{equation}

\section{Lie Group method for Fractional diffusion equation}
With the proposed fractional method of characteristics, now we can
consider a fractional Lie Group method for the fractional diffusion
equation, which are the generalizations of the classical diffusion
equations treating the super-diffusive flow processes. These
equations arise in continuous-time random walks, modeling of
anomalous diffusive and sub-diffusive systems, unification of
diffusion and wave propagation phenomenon [18 - 23].

We assume the one-parameter Lie group of transformations in
\begin{math}{\rm{(}}x,\;t,\;u)\end{math} given by
\begin{equation}
\begin{array}{l}
\frac{{\tilde x^{^\beta  } }}{{\Gamma (1 + \beta )}} =
\frac{{x^{^\beta  } }}{{\Gamma (1 + \beta )}} + \varepsilon \xi
(x,t,u)
+ O(\varepsilon ), \\
\frac{{\tilde t^{^\alpha  } }}{{\Gamma (1 + \alpha )}} =
\frac{{t^\alpha  }}{{\Gamma (1 + \alpha )}} + \varepsilon \tau
(x,t,u) + O(\varepsilon ),\\
\tilde u = u + \varepsilon \phi (x,t,u) + O(\varepsilon {\rm{),}} \\
\end{array}
\label{eq14}
\end{equation}
where \begin{math}\varepsilon \end{math} is the group parameter.

We start from the set of fractional vector fields instead of using
the one of integer order [9 - 11]
\begin{equation}
V = \xi (x,t,u)D^\beta  _x  + \tau (x,t,u)D^\alpha  _t  + \phi
(x,t,u)D_u . \label{eq15}
\end{equation}

The fractional second order prolongation \begin{math}Pr^{(2\beta )}
V\end{math} of the infinitesimal generators can be represented as
\begin{equation}
Pr^{(2\beta )} V = V + \phi ^{[t]} \frac{{\partial \phi }}{{\partial
D_t ^\alpha  u}} + \phi ^{[x]} \frac{{\partial \phi }}{{\partial D_x
^\beta  u}} + \phi ^{[tt]} \frac{{\partial \phi }}{{\partial D_t
^{2\alpha } u}} + \phi ^{[xx]} \frac{{\partial \phi }}{{\partial D_x
^{2\beta } u}} + \phi ^{[xt]} \frac{{\partial \phi }}{{\partial D_x
^\beta  D_t ^\alpha  u}}. \label{eq16}
\end{equation}

As a result, we can have
\begin{equation}
Pr^{(2\beta )} V(\Delta [u]) = 0, \label{eq17}
\end{equation}
where $\Delta [u] = \frac{{\partial ^\alpha u(x,t)}}{{\partial
t^\alpha  }} - \frac{{\partial ^{2\beta } u(x,t)}}{{\partial
x^{2\beta } }}.$

Eq. (19) can be rewritten in the form
\begin{equation}
\left. {(\phi ^{[t]}  - \phi ^{[xx]} )} \right|_{\Delta [u] = 0}  =
0. \label{eq18}
\end{equation}

The generalized prolongation vector fields are defined as
\begin{equation}\phi ^{[t]}  = D_t ^\alpha  \phi  - (D_t ^\alpha  \xi )D_x
^\beta  u - (D_t ^\alpha  \tau )D_t ^\alpha  u,\end{equation}
\begin{equation}
\phi ^{[x]}  = D_x ^\beta  \phi  - (D_x \xi ^\beta  )D_x ^\beta  u -
(D_x ^\beta  \tau )D_t ^\alpha  u, \label{eq19}
\end{equation}
\begin{equation}
\phi ^{[xx]}  = D_x ^{2\beta } \phi  - 2(D_x ^\beta  \xi )D_x
^{2\beta } u - (D_x ^{2\beta } \xi )D_x ^\beta  u - 2(D_x ^\beta
\tau )D_x ^\beta  D_t ^\alpha  u - (D_x ^{2\beta } \tau )D_t ^\alpha
u_t.
\end{equation}

Substituting Eqs. (21)-(23) into Eq. (20) and setting the
coefficients to zero, we can obtain some line fractional equations
from which we can derive
\begin{center}
\begin{displaymath}
\begin{array}{l}
\xi (x,t,u) = c_1  + c_4 \frac{{x^\beta  }}{{\Gamma (1 + \beta )}} +
2c_5 \frac{{t^\alpha  }}{{\Gamma (1 + \alpha )}} + 4c_6
\frac{{x^\beta
t^\alpha  }}{{\Gamma (1 + \beta )\Gamma (1 + \alpha )}}, \\
\tau (x,t,u) = c_2  + 2c_4 \frac{{t^\alpha  }}{{\Gamma (1 + \alpha
)}} + 4c_6 \frac{{t^{2\alpha } }}{{\Gamma (1 + 2\alpha )}}, \\
\phi (x,t,u) = (c_3  - c_5 \frac{{x^\beta  }}{{\Gamma (1 + \beta )}}
- 2c_6 \frac{{t^\alpha  }}{{\Gamma (1 + \alpha )}} - c_6
\frac{{x^{2\beta } }}{{\Gamma (1 + 2\beta )}})u + a(x,t), \\
\end{array}
\end{displaymath}
\end{center}
where $c_i (i = 0...6)$ are real constants and the function $a(x,t)$
satisfies
\begin{equation}
\frac{{\partial ^\alpha  \mathop a\limits^{} (x,t)}}{{\partial
t^\alpha  }} = \frac{{\partial ^{2\beta } \mathop a\limits^{}
(x,t)}}{{\partial x^{2\beta } }},\;\;0 < \alpha  \le 1,\;0 < \beta
\le 1.
\end{equation}

It is easy to check that the two vector fields \begin{math}\{ V_1
,V_2 ,V_3 ,V_4 ,V_5 ,V_s \} \end{math} are closed under the Lie
bracket. Thus, a basis for the Lie algebra is \begin{math}\{ V_1
,V_2 ,V_3 ,V_4 ,V_5 \}, \end{math} which consists of the
four-dimensional sub-algebra
\begin{math}\{ V_1 ,V_2 ,V_3 ,V_4 \} \end{math}
\begin{center}
\begin{displaymath}
\begin{array}{l}
v_1  = \frac{{\partial ^\beta  }}{{\partial x^\beta  }},\;\;v_2  =
\frac{{\partial ^\alpha  }}{{\partial t^\alpha  }},\;\;v_3  =
\frac{\partial }{{\partial u}},\;\;v_4  = \frac{{x^\beta  }}{{\Gamma
(1 + \beta )}}\frac{{\partial ^\beta  }}{{\partial x^\beta  }} +
\frac{{2t^\alpha  }}{{\Gamma (1 + \alpha )}}\frac{{\partial
^\alpha  }}{{\partial t^\alpha  }}, \\
v_5  = \frac{{2t^\alpha  }}{{\Gamma (1 + \alpha )}}\frac{{\partial
^\beta  }}{{\partial x^\beta  }} - \frac{{ux^\beta  }}{{\Gamma (1 +
\beta )}}\frac{\partial }{{\partial u}},\;\; \\
v_6  = \frac{{4t^\alpha  }}{{\Gamma (1 + \alpha )}}\frac{{x^\beta
}}{{\Gamma (1 + \beta )}}\frac{{\partial ^\beta  }}{{\partial
x^\beta }} + \frac{{4t^{2\alpha } }}{{\Gamma (1 + 2\alpha
)}}\frac{{\partial ^\alpha  }}{{\partial t^\alpha  }} -
(\frac{{x^{2\beta } }}{{\Gamma (1 + 2\beta )}} + \frac{{2t^\alpha
}}{{\Gamma (1 + \alpha
)}})u\frac{\partial }{{\partial u}}{\rm{,}} \\
\end{array}
\end{displaymath}
\end{center}
and one infinite-dimensional sub-algebra
\begin{equation}
v_7  = a(x,t)\frac{\partial }{{\partial u}}.
\end{equation}

Assume \begin{math}u = f(\frac{{x^\beta  }}{{\Gamma (1 + \alpha
)}},\;\frac{{t^\alpha  }}{{\Gamma (1 + \beta )}})\end{math} is an
exact solution of Eq. (4). Then with the proposed fractional method
of characteristics, solving the above symmetry equations, we can
derive
\begin{displaymath}
\begin{array}{l}
u^{{\rm{(1)}}}  = f(\frac{{x^\beta  }}{{\Gamma (1 + \alpha )}} -
\varepsilon ,\frac{{t^\alpha  }}{{\Gamma (1 + \alpha )}}){\rm{,}} \\
u^{{\rm{(2)}}}  = f(\frac{{x^\beta  }}{{\Gamma (1 + \beta
)}},\frac{{t^\alpha  }}{{\Gamma (1 + \alpha )}} - \varepsilon
){\rm{,}}
\\
u^{{\rm{(3)}}}  = e^\varepsilon  f(\frac{{x^\beta  }}{{\Gamma (1 +
\beta )}},\;\frac{{t^\alpha  }}{{\Gamma (1 + \alpha )}}){\rm{,}} \\
u^{{\rm{(4)}}}  = f(\frac{{x^\beta  }}{{\Gamma (1 + \beta
)}}e^{-\varepsilon}  ,\frac{{t^\alpha  }}{{\Gamma (1 + \alpha )}}e^{
-
2\varepsilon } {\rm{),}} \\
u^{{\rm{(5)}}}  = e^{\frac{{t^\alpha  \varepsilon ^2 }}{{\Gamma (1 +
\alpha )}} - \frac{{x^\beta  \varepsilon }}{{\Gamma (1 + \beta )}}}
f(\frac{{x^\beta  }}{{\Gamma (1 + \beta )}} - 2\varepsilon
\frac{{t^\alpha  }}{{\Gamma (1 + \alpha )}},\frac{{t^\alpha
}}{{\Gamma
(1 + \alpha )}}){\rm{,}} \\
u^{{\rm{(6)}}}  = \frac{1}{{\sqrt {1 + 4\varepsilon \frac{{t^\alpha
}}{{\Gamma (1 + \alpha )}}} }}e^{\frac{{ - x^{2\beta } \varepsilon
\Gamma (1 + \alpha )}}{{\Gamma (1 + 2\beta )\Gamma (1 + \alpha ) +
4\varepsilon t^\alpha  \Gamma (1 + 2\beta )}}}  \\
\times f(\frac{{\Gamma (1 + \alpha )x^\beta  }}{{\Gamma (1 + \beta
)\Gamma (1 + \alpha ) + 4\varepsilon \Gamma (1 + \alpha )x^\beta
}},\frac{{t^\alpha  }}{{\Gamma (1 + \beta ) + 4\varepsilon \Gamma (1
+
\alpha )t^\alpha  }}){\rm{,}} \\
\\
u^{{\rm{(7)}}}  = f(\frac{{x^\beta  }}{{\Gamma (1 + \alpha
)}},\frac{{t^\alpha  }}{{\Gamma (1 + \alpha )}}) + \varepsilon
a(x,t){\rm{,}} \\
\end{array}
\end{displaymath}
which are all the classification of solutions of Eq. (4).

Take the solution \begin{math}u^{{\rm{(5)}}} \end{math} as an
example,

\begin{equation}
u^{{\rm{(5)}}}= e^{\frac{{t^\alpha \varepsilon ^2 }}{{\Gamma (1 +
\alpha )}} - \frac{{x^{\beta}  \varepsilon }}{{\Gamma (1 + \beta
)}}} f(\frac{{x^\beta  }}{{\Gamma (1 + \beta )}} - 2\varepsilon
\frac{{t^\alpha  }}{{\Gamma (1 + \alpha )}},~\frac{{t^\alpha
}}{{\Gamma (1 + \alpha )}}). \label{eq20}
\end{equation}

Assume \begin{math}f(\frac{{x^\beta  }}{{\Gamma (1 + \beta )}} -
2\varepsilon \frac{{t^\alpha  }}{{\Gamma (1 + \alpha
)}},~\frac{{t^\alpha  }}{{\Gamma (1 + \alpha )}}) =
c,~\end{math}which can be set as the initial value of Eq. (4). Now
we can check that \begin{math}u_1 ^{{\rm{(5)}}}  =
ce^{\frac{{t^\beta \varepsilon ^2 }}{{\Gamma (1 + \beta )}} -
\frac{{x^\alpha \varepsilon }}{{\Gamma (1 + \alpha )}}} \end{math}
is one of the exact solutions. If we make
\begin{math}f(\frac{{x^\beta }}{{\Gamma (1 + \alpha
)}},\frac{{t^\alpha  }}{{\Gamma (1 + \alpha )}}) = u_1 ^{{\rm{(5)}}}
= ce^{\frac{{x^\beta  \varepsilon ^2 }}{{\Gamma (1 + \beta )}} -
\frac{{t^\alpha  \varepsilon }}{{\Gamma (1 + \alpha )}}}
\end{math}, we can derive a new iteration solution \begin{math}u_2
^{{\rm{(5)}}} \end{math}. As a result, by similar manipulations, we
can give \begin{math}u_{_3 } ^{{\rm{(5)}}}
\end{math}\ldots{}\begin{math}u_n ^{{\rm{(5)}}} \end{math} which are
new exact solutions of Eq. (4).

\section{Conclusions}
Fractional differential equations have caught considerable attention
due to their various applications in real physical problems.
However, there is no systematic method to derive the exact solution.
Now, the problem is partly solved in this paper.

Another problem may arise: can the Lie group method be extended to
fractional differential equations of fractional order 0 $\thicksim$
2? We will discuss such work in future.

\clearpage

\end{document}